\documentstyle[11pt,paspconf,epsf]{article}

\def\ion#1#2{\ifmmode \mbox{{\rm #1}}\,\mbox{{\sc #2}} 
        \else {\rm #1}$\,${\sc #2}
        \fi}

\def\unitspace{\,}                      

\def\un#1{\ifmmode \unitspace{\rm #1} 
          \else $\unitspace$#1
          \fi}
\def\pun#1#2{\ifmmode \unitspace\mbox{\rm #1}^{#2} 
             \else $\unitspace$#1$^{#2}$
             \fi}

\def\kms{\un{km}\pun{s}{-1}}          
\def\Lsun{\ifmmode \un{L}_{\odot}     
          \else $\un{L}_{\odot}$
          \fi}
\def\mum{\ifmmode \unitspace\mu\mbox{\rm m} 
         \else $\unitspace\mu$m
         \fi}

\def\sou#1#2{\relax                       
             \ifmmode {\rm #1}\,{\rm #2}  
             \else #1$\,$#2
             \fi}

\def\simlt{\mathrel{\hbox{\rlap{\hbox{\lower4pt\hbox{$\sim$}}}\hbox{$<$}}}}
\def\simgt{\mathrel{\hbox{\rlap{\hbox{\lower4pt\hbox{$\sim$}}}\hbox{$>$}}}}

\begin{document}

\makeatletter
\renewcommand{\@makefnmark}{\mbox{\ }}
\makeatother

\renewcommand{\thefootnote}{}

\footnote{to appear in {\it
``Highly Redshifted Radio Lines''}, eds.\ C.\ L.\ Carilli,
S.\ J.\ E.\ Radford, K.\ Menten,  \& G.\ Langston, Astronomical
Society of the Pacific, San Francisco}

\title{Observations and diagnostic use of highly redshifted
fine structure lines}

\author{Paul P.\ van der Werf}
\affil{Leiden Observatory, P.O.\ Box 9513,
NL - 2300 RA Leiden, The Netherlands (pvdwerf@strw.leidenuniv.nl
)}

{\footnotesize
\begin{abstract}
The diagnostic use and detectability of luminous fine structure lines
from high redshift galaxies is reviewed in the light of results from
COBE concerning the Milky Way and from ISO on low redshift galaxies.
At the highest luminosities ($L>10^{12}\Lsun$)
the [$\ion{C}{ii}$] $158\mum$ line is somewhat less luminous with
respect to the bolometric luminosity than for lower luminosity
objects. Thus, surveys for this line must emphasize depth. The
[$\ion{C}{ii}$] line will be the principal spectroscopic
probe of the deep universe for the MMA and FIRST\null. A deep search
for [$\ion{C}{ii}$] $158\mum$ emission from the dusty $z=4.693$ quasar
$\sou{BR}{1202{-}0725}$ is presented. The resulting $3\sigma$ upper
limit implies that for this object $L_{\rm [CII]}/L_{\rm
FIR}<0.0006\%$, a highly significant result indicating that distant
luminous objects may represent a natural extension towards higher
luminosities of the ultraluminous infrared galaxies at low redshift.
\end{abstract}
}


\section{Introduction}

Warm, neutral interstellar gas cools mainly through emission in low
excitation fine structure lines, principally the [$\ion{C}{ii}$]
$158\mum$ and [$\ion{O}{i}$] $63\mum$ lines. Star forming galaxies
therefore emit copious amounts of radiation in these lines, as shown
observationally by observations with the KAO
(e.g., Stacey et al.\ 1991) and the ISO satelite (e.g., Malhotra et
al.\ 1997). The brightest of these lines exceed the brightest CO
lines by one to two orders of magnitude in luminosity.  These lines,
which lie mostly in the far-infrared (FIR) region, are therefore
expected to be detectable out to very large distances where they are
redshifted into the submillimeter region, and their detection has long
been recognized as a key aim of extragalactic submillimeter spectroscopy.
The discovery of luminous CO emission lines in a number of high-$z$
objects has intensified searches for fine structure lines, which, as
principal coolants, carry substantial diagnostic information on the
heating sources involved, which may be immense, optically obscured
starbursts.

\begin{table}
\caption{Luminosities of fine structure lines in the Milky Way. 
$L_{\rm FIR}=10^{10}\Lsun$ is the far-infrared luminosity of the Milky Way.}
\label{tab.MWfluxes}
\begin{center}
\begin{tabular}{c@{\qquad}c@{\qquad}r@{\qquad}c@{\qquad}l}
\tableline\\[-3mm]
Line & $\lambda_0$ & $\nu_0$ & $L_{\rm line}$ & $L_{\rm line}/L_{\rm
FIR}$\\
     & [$\mum$]    & [GHz]   & [$\Lsun$]      &            \\[1mm]
\tableline\\[-3mm]
\protect[$\protect\ion{C}{ii}$] & 158 & 1900 & $5\cdot10^7$ & 0.5\% \\
\protect[$\protect\ion{O}{i}$]  &  63 & 4745 & $2\cdot10^7$ & 0.2\% \\
\protect[$\protect\ion{N}{ii}$] & 122 & 2459 & $8\cdot10^6$ & 0.08\% \\
\protect[$\protect\ion{N}{ii}$] & 204 & 1461 & $5\cdot10^6$ & 0.05\% \\
\protect[$\protect\ion{C}{i}$]  & 370 &  809 & $3\cdot10^5$ & 0.003\% \\
\protect[$\protect\ion{C}{i}$]  & 609 &  492 & $2\cdot10^5$ & 0.002\% \\[1mm]
\tableline
\tableline
\end{tabular}
\end{center}
\end{table}

In this review, I first discuss the current state of knowledge on the
properties of these lines in the Milky Way and local galaxies. I
briefly address the detectability and diagnostic use of these lines at
high $z$ with future instrumentation.  Finally I present the deepest
search so far for redshifted [$\ion{C}{ii}$] $158\mum$ emission,
which provides for the first time a highly significant upper limit, and
may provide a link between dusty, luminous high-$z$ objects and
ultraluminous infrared galaxies (ULIRGs) in the local universe.

\begin{figure}
\plotfiddle{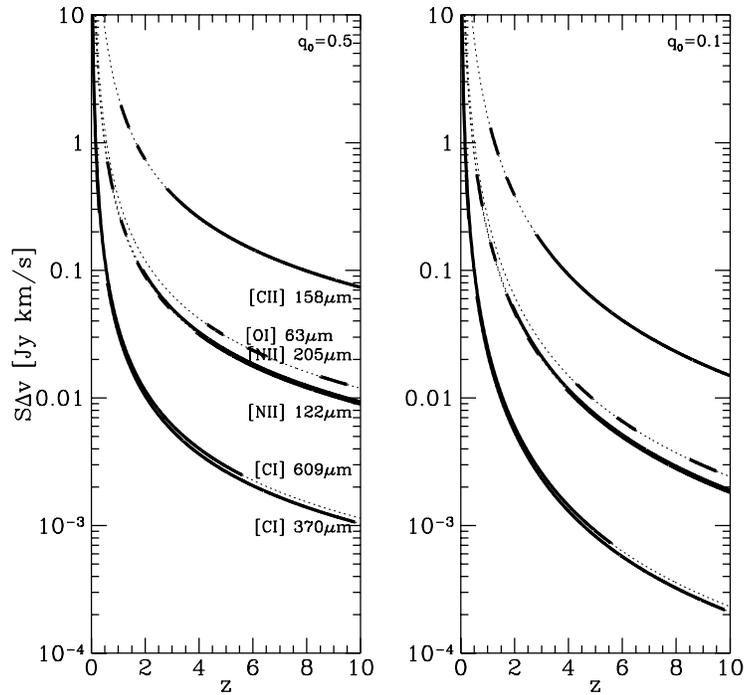}{10cm}{0}{50}{50}{-160}{-72}
\caption{Expected integrated emission line fluxes of the Milky Way as
a function of redshift for $H_0=75\kms\,{\rm Mpc}^{-1}$ and $q_0=0.5$
(left panel) and $q_0=0.1$ (right panel), calculated as described in
Van der Werf and Israel (1996). Drawn lines indicate
atmospheric windows accessible from the ground: $75-500\,$GHz
(some narrow inaccessible frequency ranges have been ignored), 
$630-700\,$GHz, and $800-900\,$GHz; dashed lines denote regions
inaccesible due to atmospheric opacity.}
\label{fig.MWfluxes}
\end{figure}

\section{The Milky Way at low and high redshift}

The first accurate determinations of the global luminosities of luminous
fine structure lines in the Milky Way have been provided by 
the FIRAS experiment on the COBE satelite (Wright et al.\ 1991). These
luminosities are summarized in Table~\ref{tab.MWfluxes}, together with
a much more uncertain luminosity of [$\ion{O}{i}$] $63\mum$, which
has not been measured by FIRAS, but estimated from the possible FIRAS
detection of [$\ion{O}{i}$] $145\mum$, using a [$\ion{O}{i}$]
$63\mum/145\mum$ ratio of 20, and which agrees with the estimate by
Stacey (1989). 

Given these luminosities, it is easy to calculate the apparent
integrated emission line fluxes for the Milky Way as a function of
redshift. The expected fluxes are presented
in Figure~\ref{fig.MWfluxes}.

It is clear that the [$\ion{C}{ii}$] line is going to be the most
important spectroscopic probe of the deep universe in this frequency
region, once sufficiently sensitive facilities become available. Using
currently expected sensitivities for the MMA (assuming a total
collecting area of $7000\,$m$^2$), for $q_0=0.5$ the Milky Way will be
detectable as a point source for the MMA in an 8 hour synthesis at
{\it any\/} redshift where the line is shifted into a transparent
atmospheric window. Mapping will be more difficult (because the
emission is divided into a number of beams, and because of
cosmological surface brightness dimming), but should still be
routinely possible in objects somewhat more luminous than the Milky
Way, or in longer integrations. [$\ion{N}{ii}$] and in particular
[$\ion{O}{i}$] measurements will be limited to more luminous objects
in specific redshift intervals. The [$\ion{C}{i}$] lines from the
Milky Way will only be detectable out to $z\sim1$. The
[$\ion{C}{ii}$] line will not be accessible to the MMA at $z<1$, but
this redshift range will be covered by the heterodyne instrument
HIFI on the FIRST satelite.

\section{Diagnostic use of cooling lines}

Since the low excitation fine structure lines 
are major cooling lines, their fluxes
should scale to zeroth order with the heating rate, and hence with the
FIR luminosity $L_{\rm FIR}$. KAO observations (Crawford et al.\ 1985; 
Stacey et al.\ 1991) confirmed this behaviour for the [$\ion{C}{ii}$]
line, which carries typically $0.5\%$ of the FIR luminosity for 
$L_{\rm FIR}\simlt 10^{11}\Lsun$. Detection of these lines at high
redshift will provide
powerful diagnostics.
\begin{enumerate}
\item The [$\ion{C}{ii}$] line originates at the UV-exposed surfaces
of molecular clouds (photon-dominated regions or PDRs). 
It becomes a particular powerful diagnostic in
combination with CO lines, since the [$\ion{C}{ii}$]/CO line ratio is
independent of beam filling factor, and thus provides an
extinction-free measurement of the UV luminosity per unit molecular
mass, i.e., of the star forming efficiency. Combining this result with
a star formation rate and dust mass determined from $L_{\rm FIR}$ provides a 
characterization of the star forming properties of distant galaxies
(Crawford et al.\ 1985; Stacey et al.\ 1991).
\item The [$\ion{O}{i}$] line also arises in PDRs, and, combined with
[$\ion{C}{ii}$], provides an independent measure of the UV field
strength.
\item The [$\ion{C}{i}$] lines also arise at molecular cloud surfaces,
and may be used to trace the mass of UV-exposed gas.
\item The [$\ion{N}{ii}$] lines, with a formation potential of only
$14.53\,$eV, and low critical density and upper level temperature, are
ideal probes of the general low excitation ionized medium.
\end{enumerate}

\section{Lessons from local galaxies}

The recent discovery of high redshift objects with FIR luminosities
up to $10^{14}\Lsun$ and the
detection of CO emission in several of these 
have led to the suggestion of immense, optically
obscured starbursts at high redshift. Application of the simple
scaling of $L_{\rm [CII]}$ with $L_{\rm FIR}$ suggested in the
preceding section results in [$\ion{C}{ii}$] fluxes for these luminous
objects that should already be observable with present instrumentation.
However, searches for redshifted [$\ion{C}{ii}$] $158\mum$ (Isaak et
al.\ 1994; Ivison et al.\ 1998) and [$\ion{N}{ii}$] $205\mum$ emission
(Ivison and Harrison 1996) have not yielded any detections and the
only reliably detected fine structure line at high redshift so far is
the [$\ion{C}{i}$] $609\mum$ detection of the gravitationally lensed
``Cloverleaf quasar'' (Barvainis et al.\ 1997).

Recent FIR spectroscopy of low redshift galaxies with the ISO Long
Wavelength Spectrograph (LWS) has shed new light on
these results. As shown by Malhotra et al.\ (1997), the ratio
$L_{\rm [CII]}/L_{\rm FIR}$ falls dramatically below 0.5\% for
galaxies with high star formation rates, which typically have
$L_{\rm [CII]}/L_{\rm FIR}\sim0.05\%$. The most outstanding
example of this phenomenon is the prototypical ULIRG Arp$\,220$, where
the [$\ion{C}{ii}$] line is strongly suppressed with respect to the
continuum (Fischer et al.\ 1998).  While the origin of this effect is
still debated (optically thick [$\ion{C}{ii}$] emission, dust
absorption at $158\mum$, and suppressed [$\ion{C}{ii}$] emission in
dense PDRs may all play a role), it is evident that at the
luminosities characterizing ULIRGs the [$\ion{C}{ii}$] luminosity does
not scale with $L_{\rm FIR}$ anymore, but is suppressed by about a
factor of 10 with respect to the lower luminosity scaling (Luhman et
al.\ 1998). The behaviour of other fine structure lines is even more
complicated: in Arp\,220, [$\ion{O}{i}$] $63\mum$ (and many lines of
OH, H$_2$O, NH$_3$ and CH) is in {\it absorption\/} (Fischer et al.\
1998)! In the somewhat less luminous starburst galaxy NGC\,3690, both
[$\ion{C}{ii}$] and [$\ion{O}{i}$] are bright, and neither Arp\,220
nor NGC\,3690 shows [$\ion{N}{ii}$], but the latter object has
[$\ion{N}{iii}$] emission (Fischer et al.\ 1998). In contrast,
the modest nearby starburst galaxies M82 and NGC\,253 show both
[$\ion{N}{ii}$] and [$\ion{N}{iii}$] emission (Hur et al.\ 1996).

While it will be some time before these results can be placed in
their proper physical context, a number of phenomenological
conclusions relevant to high-$z$
observations of these lines can already be drawn.
\begin{enumerate}
\item In objects with luminosities up to $\sim10^{12}\Lsun$, the
[$\ion{C}{ii}$] $158\mum$ line carries about 0.5\% of $L_{\rm
FIR}$. In more luminous objects this fraction decreases to about
$0.05\%$.
\item On the other hand, in low-metallicity objects such as the
Magellanic clouds, the [$\ion{C}{ii}$] line carries about 1\% of $L_{\rm
FIR}$ and this fraction may increase to about 3\% in individual star
forming regions in these objects (Israel et al.\ 1996). 
This behaviour results from the
easier photodissociation of CO in low metallicity galaxies, and may
positively affect detection rates for high-$z$ objects.
\item The nitrogen lines are affected by excitation conditions, and in
the absence of detailed further information, upper limits will be extremely
difficult to interpret. In addition, at high
redshift the strengths of the nitrogen lines may be affected by
abundance effects.
\item \protect{[$\ion{O}{i}$]} $63\mum$ may be affected by radiative transfer
effects. Consequently,
detections of this line will be difficult to interpret.
\end{enumerate}

It is important to note that the use of [$\ion{C}{ii}$] emission as a
cosmological probe for the MMA is not compromised by these results, since
the MMA will be able to probe sufficiently far to the faint end of the
luminosity function. Only for ultraluminous objects flux predictions
have to be lowered in the light of the ISO results. These results also
imply that blind surveys for
extragalactic [$\ion{C}{ii}$] emission (see e.g., Stark 1997 for
predictions of detection rates) should emphasize survey depth
rather than survey volume, until the luminosity function is probed to
sufficiently faint levels at the highest redshift of interest. 

\begin{figure}[ht]
\plotfiddle{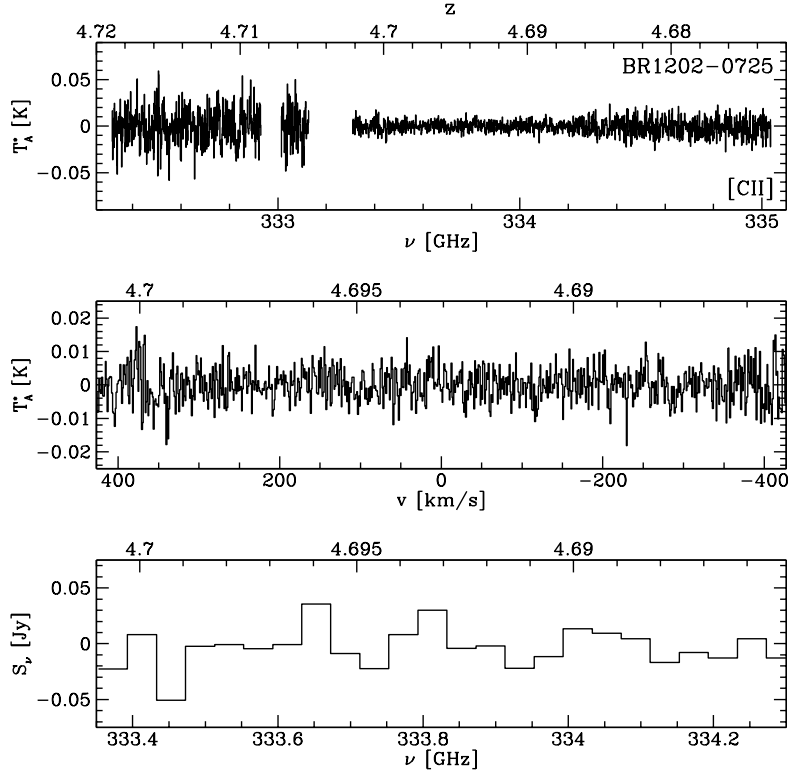}{9.5cm}{0}{54}{54}{-170}{-100}
\caption{Final combined spectra of [$\ion{C}{ii}$] $158\mum$ emission
from $\sou{BR}{1202{-}0725}$ at $z=4.693$. The top panel shows the
full spectral range covered, at a frequency resolution of 1.513\,MHz
(velocity resolution $1.36\kms$); for reasons of presentation, the
most noisy sections have not been plotted. The central panel shows a
blow-up of the most sensitive section of the spectrum, centered on the
expected frequency of the line, which is taken as the zeropoint of the
indicated velocity scale. The bottom panel shows the same section,
smoothed to 40\,MHz frequency
resolution ($36\kms$ velocity resolution), and converted into flux
density units.}
\label{fig.spectra}
\end{figure}

\section{A deep search for [C\,II] emission from BR\,$\mathbf{1202{-}0725}$ at
$\mathbf{z=4.69}$}

Even with the more pessimistic flux predictions discussed in
Section~4, the most luminous high-$z$ objects might be
detectable in redshifted [$\ion{C}{ii}$] $158\mum$ emission in a long
integration with presently available instrumentation. 
We have therefore obtained a very deep spectrum of the
dusty $z=4.69$ radio-quiet QSO $\sou{BR}{1202{-}0725}$ (Van der Werf
and Israel, in preparation). This object shows strong submillimeter
dust emission (McMahon et al.\ 1994; Isaak et al.\ 1994), with $L_{\rm
FIR}\sim10^{14}\Lsun$ ($H_0=75\kms\,{\rm Mpc}^{-1}$, $q_0=0.1$ is
adopted here and in the remainder of this paper). Luminous CO emission
was detected by Ohta et al.\ (1996) and Omont et al.\ (1996) and
provided an accurate redshift of 4.693, so that the [$\ion{C}{ii}$]
$158\mum$ line is redshifted to a frequency of 333.8375\,GHz, where
the atmosphere is transparant. The JCMT was used to search for this
line during several observing runs between December 1993 and April
1997, using first the B3(i) single-channel SIS receiver, and later the
B3 dual-channel SIS receiver. The DAS autocorrelator spectrograph was
used as backend. Since at the start of this project the CO detection
(and hence the precise redshift) was not available yet, a range
of frequencies had to be scanned. This situation underlines the need
for very wideband systems (e.g., Isaak, these proceedings) 
for future observations of fine structure
lines from distant objects. The observations
were done with double beam switching, so that the source appeared
alternately in the signal and reference beams of the chopping
secondary mirror. The subbands of the DAS were merged using SPECX, and
further processing was done in CLASS\null. All scans were visually
inspected and those with curved baselines or zeropoint offsets were
discarded, leaving a total integration time of about 50~hours on
source. In order to safeguard against spurious broad features,
horizontal baselines were subtracted from all individual frequency
settings before merging these into a final composite spectrum (cf.,
Ivison et al.\ 1998). The resulting spectrum is shown in
Figure~\ref{fig.spectra}.

\begin{table}
\caption{Empirical luminosity ratios for various types of objects.}
\label{tab.ratios}
\begin{center}
\begin{tabular}{lcc}
\tableline\\[-3mm]
& $L_{\rm [CII]}/L_{\rm CO(1-0)}$ & $L_{\rm [CII]}/L_{\rm FIR}$ \\[1mm]
\tableline\\[-3mm]
Milky Way          &  1400 & 0.5\%  \\
Starburst galaxies &  4100 & 0.5\%  \\
ULIRGs             &  1700 & 0.05\% \\
Orion region       &  7000 & 0.1\%  \\
LMC                & 23000 & 1\%    \\
30~Dor             & 77000 & 2.5\%  \\
\tableline
\tableline
\end{tabular}
\end{center}
\end{table}

The [$\ion{C}{ii}$] line is not detected. Assuming a Gaussian line with a
FWHM of $220\kms$ (as measured by Ohta et al.\ 1996 
for the CO $J=5{\to}4$ line), the $3\sigma$ upper limit to the
integrated flux 
is $9.5\,{\rm Jy}\kms$, so that $L_{\rm [CII]}<6.3\cdot10^9\Lsun$, and
$L_{\rm [CII]}/L_{\rm FIR}<0.006\%$. Another useful diagnostic ratio,
with CO $J=1{\to}0$, can be estimated from the CO $J=5{\to}4$
measurements, noting that the CO excitation properties are likely
similar to those in the Cloverleaf quasar (Omont et al.\ 1996), and
using the estimate by Barvainis et al.\ (1997) for the CO
$J=1{\to}0/J=5{\to}4$ ratio
of the Cloverleaf. The resulting $3\sigma$ upper limit is $L_{\rm
[CII]}/L_{\rm CO(1-0)}<4000$.

These results can be put in perspective using
Table~\ref{tab.ratios}, which is based on data from Stacey et al.\ (1991),
Nakagawa et al.\ (1993), Solomon et al.\ (1997), and Luhman et al.\
(1998). It is evident that the low $L_{\rm [CII]}/L_{\rm FIR}$ ratio,
which is at least a factor of 8 smaller than in local ULIRGs, 
is a highly significant result. The low $L_{\rm [CII]}/L_{\rm CO(1-0)}$
ratio is consistent with this result, without providing an
additional constraint. Table~\ref{tab.ratios} indicates that we should
go at least a factor of 3 deeper before the same $L_{\rm [CII]}/L_{\rm
CO(1-0)}$ ratio is reached as in local ULIRGs. Such sensitivities are
beyond the reach of presently available instrumentation, but will be
available with the MMA\null. These results suggest that distant, dusty
hyperluminous objects such as $\sou{BR}{1202{-}0725}$ represent a natural
extension towards higher luminosities of the local population of
ULIRGs. Hence, the study of local ULIRGs will also shed light on the
role and properties of distant hyperluminous objects, whether they are
powered by active nuclei or immense bursts of star formation.

\section{Conclusions}

Redshifted fine structure lines form powerful diagnostics of the
physical conditions in the neutral interstellar gas in distant
galaxies. The [$\ion{C}{ii}$] $158\mum$ will form the principal
spectroscopic probe of the deep universe for the MMA, which (for
$q_0=0.5$) will be
able to detect this line from galaxies down to Milky Way luminosities at
every redshift where the line is shifted into a transparent
atmospheric window. A deep search for redshifted [$\ion{C}{ii}$]
$158\mum$ emission from the dusty $z=4.693$ quasar
$\sou{BR}{1202{-}0725}$ has resulted in highly significant $3\sigma$
upper limit $L_{\rm [CII]}/L_{\rm FIR}<0.0006\%$, at least factor 8
lower than ULIRGs at low $z$, and a factor of 80 lower than
luminous starburst galaxies, suggesting that hyperluminous,
dusty high redshift objects such as $\sou{BR}{1202{-}0725}$ represent
an extension towards higher luminosities of the population of ULIRGs
in the local universe.

\acknowledgments

It is a pleasure to thank Remo Tilanus for taking most of the JCMT
data presented in this paper.

\end{document}